\documentclass[aps,pre,preprint,amsmath,amssymb]{revtex4-1}

\usepackage[english]{babel}
\usepackage[utf8]{inputenc}
\usepackage[T1]{fontenc} 		
\usepackage{times}
\usepackage{color}			
\usepackage{graphicx}
\usepackage{xspace}
\usepackage{SIunits}
\usepackage{comment}			

\newcommand{\FIG}[3]{
\begin{figure}[htbp]
\begin{center}
\includegraphics[width=\columnwidth]{#2}
\end{center}
\caption{#3\label{fig:#1}}
\end{figure}}

\newcommand{\SFIG}[3]{
\begin{figure}[htbp]
\begin{center}
\includegraphics[width=0.60\columnwidth]{#2}
\end{center}
\caption{#3\label{fig:#1}}
\end{figure}}

\def\eq#1{Eq.~\eqref{eq:#1}\xspace}

\def\fig#1{Fig.~\ref{fig:#1}\xspace}

\def\nm{\nano\meter\xspace}
\def\pN{\pico\newton\xspace}
\def\pNpernm{\pN\nm^{-1}\xspace}

\def\hz{\second^{-1}\xspace}

\def\const{\mathrm{const}\xspace}

\def\kbT{k_{\mathrm{B}}T\xspace}

\def\ee{\mathrm{e}\xspace}
\def\diff{\mathrm{d}\xspace}
\newcommand{\Kronecker}[2]{\delta_{#1,#2}\xspace}
\newcommand{\dd}[2]{\frac{\diff}{\diff #2} #1 \xspace}

\def\Myo{Myosin {II}\xspace}
\def\myo{myosin {II}\xspace}
\def\atp{ATP\xspace}
\def\adp{ADP\xspace}
\def\pho{P$_{\textrm{i}}$\xspace}

\def\pcm{PCM\xspace}
\def\lte{LTE\xspace}

\def\ub{\emph{ub}\xspace}
\def\wb{\emph{wb}\xspace}
\def\pps{\emph{pps}\xspace}

\def\pe{$(+)$\xspace}
\def\me{$(-)$\xspace}
\def\pme{$(\pm)$\xspace}

\def\ken{k_{10}\xspace}
\def\kzn{k_{20}\xspace}
\def\kez{k_{12}\xspace}
\def\kze{k_{21}\xspace}
\def\kne{k_{01}\xspace}

\def\kzns{{\tilde k}_{20}\xspace}

\def\Fdelta{F_{0}\xspace}

\def\Fbar{{\bar F}\xspace}

\def\rpm{r_{\pm}\xspace}
\def\gpm{g_{\pm}\xspace}
\def\rplu{r_{+}\xspace}
\def\rmin{r_{-}\xspace}
\def\gplu{g_{+}\xspace}
\def\gmin{g_{-}\xspace}

\def\Epp{E_{\mathrm{pp}}\xspace}
\def\Eext{E_{\mathrm{ext}}\xspace}
\def\Eelp{E_{\mathrm{el,+}}\xspace}
\def\Eelm{E_{\mathrm{el,-}}\xspace}
\def\Eelpm{E_{\mathrm{el,\pm}}\xspace}

\def\kl{k_{\mathrm{m}}\xspace}
\def\Kp{K_{+}\xspace}
\def\Km{K_{-}\xspace}
\def\Kpm{K_{\pm}\xspace}
\def\Kext{K_{\mathrm{ext}}\xspace}
\def\Ktot{K_{\mathrm{tot}}\xspace}

\def\zsum{z\xspace}
\def\zabs{z_{\textrm{abs}}\xspace}
\def\zpm{z_{\pm}\xspace}
\def\zmp{z_{\mp}\xspace}
\def\zp{z_{+}\xspace}
\def\zm{z_{-}\xspace}

\def\zpmbar{{\bar z}_{\pm}\xspace}
\def\xpmbar{{\bar x}_{\pm}\xspace}

\def\xpm{x_{\pm}\xspace}
\def\xmp{x_{\mp}\xspace}
\def\xp{x_{+}\xspace}
\def\xm{x_{-}\xspace}

\def\xneg{x_{\mathrm{neg}}\xspace}
\def\xpos{x_{\mathrm{pos}}\xspace}

\def\fpm{F_{\pm}\xspace}
\def\fext{F_{\rm ext}\xspace}

\def\Npm{N_{\pm}\xspace}
\def\Np{N_{+}\xspace}
\def\Nm{N_{-}\xspace}

\def\ipm{i_{\pm}\xspace}
\def\imp{i_{\mp}\xspace}
\def\ip{i_{+}\xspace}
\def\im{i_{-}\xspace}

\def\jpm{j_{\pm}\xspace}
\def\jmp{j_{\mp}\xspace}
\def\jp{j_{+}\xspace}
\def\jm{j_{-}\xspace}

\def\yp{\ip,\jp\xspace}
\def\ym{\im,\jm\xspace}
\def\I{\ip,\im\xspace}
\def\J{\jp,\jm\xspace}

\newcommand{\pwidjArgs}{p\left(\J|\I\right)\xspace}
\newcommand{\pipim}[2]{p_{#1,#2}\xspace}
\newcommand{\pipimss}[2]{p_{#1,#2}(\infty)\xspace}

\renewcommand{\leq}{\leqslant}
\renewcommand{\geq}{\geqslant}

\begin{document}

\date{\today}

\title{Stochastic dynamics and mechanosensitivity of \myo minifilaments}

\author{Philipp J.~Albert} \affiliation{BioQuant, Heidelberg University, Im Neuenheimer Feld 267, 69120 Heidelberg, Germany and Institute for Theoretical Physics, Heidelberg University, Philosophenweg 19, 69120 Heidelberg, Germany} 
\author{Thorsten Erdmann}  \affiliation{BioQuant, Heidelberg University, Im Neuenheimer Feld 267, 69120 Heidelberg, Germany and Institute for Theoretical Physics, Heidelberg University, Philosophenweg 19, 69120 Heidelberg, Germany} 
\author{Ulrich S.~Schwarz$^\ast$} \affiliation{BioQuant, Heidelberg University, Im Neuenheimer Feld 267, 69120 Heidelberg, Germany and Institute for Theoretical Physics, Heidelberg University, Philosophenweg 19, 69120 Heidelberg, Germany} 

\begin{abstract}
{Tissue cells are in a state of permanent mechanical tension that is
maintained mainly by \myo minifilaments, which are bipolar assemblies of
tens of \myo molecular motors contracting actin networks and bundles.
Here we introduce a stochastic model for \myo minifilaments as two small
\myo motor ensembles engaging in a stochastic tug-of-war. Each of the
two ensembles is described by the parallel cluster model that allows us
to use exact stochastic simulations and at the same time to keep
important molecular details of the \myo cross-bridge cycle. Our
simulation and analytical results reveal a strong dependence of \myo
minifilament dynamics on environmental stiffness that is reminiscent of
the cellular response to substrate stiffness. For small stiffness,
minifilaments form transient crosslinks exerting short spikes of force
with negligible mean. For large stiffness, minifilaments form near
permanent crosslinks exerting a mean force which hardly depends on
environmental elasticity. This functional switch arises because
dissociation after the power stroke is suppressed by force (catch
bonding) and because ensembles can no longer perform the power stroke at
large forces. Symmetric \myo minifilaments perform a random walk with an
effective diffusion constant which decreases with increasing ensemble
size, as demonstrated for rigid substrates with an analytical treatment.}\\
{$^\ast$Correspondence: Ulrich.Schwarz@bioquant.uni-heidelberg.de}
\end{abstract}

\maketitle


\section{Introduction}

Cytoskeletal molecular motors are a large class of proteins that
generate movement and force in biological cells by cycling between
states bound and unbound from a cytoskeletal filament \cite{Howard1997,
Schliwa2003}.  In general, they can be classified as processive or
non-processive motors.  Processive motors like kinesin, dynein or myosin
V have a duty ratio (fraction of time of the motor cycle spent on the
filament) close to unity and therefore are particularly suited for
persistent transport of cellular cargo, such as vesicles, small
organelles or viruses. Using small groups of processive motors increases
the walk length and the efficiency of transport compared to the single
motor \cite{holzbaur_coordination_2010}. A theoretical
treatment with a one-step master equation showed that the effective
unbinding rate decreases exponentially with the size of the motor ensemble
\cite{Klumpp2005}. Moreover groups of motors can also work
against larger load than the single motor \cite{Koster2003}. If motors
of different directionality on the substrate are attached to the same
cargo, bidirectional movement can ensue \cite{Muller2008},
as often observed in cargo transport. A similar tug-of-war
setup has been used earlier to explain mitotic spindle oscillations
\cite{grill_theory_2005}. Non-processive motors such as \myo have a duty ratio
significantly smaller than unity. Therefore, non-processive motors have
to operate in groups in order to generate appreciable levels of force.
Similar to processive motors, the duty ratio of a group of
non-processive motors increases with the size of the group and can
become large enough that the group effectively behaves like a processive
motor. This is certainly true for the sarcomeres in skeletal muscle,
where typically hundreds of \myo work together as one group. Combining
structural investigations of skeletal muscle with modeling has led to
the swinging cross-bridge model for single \myo \cite{Huxley1957,
Huxley1971}. A statistical treatment then has allowed to accurately
model the dynamics of the motor ensemble in muscle sarcomeres
\cite{Duke1999, Vilfan2003}.

Groups of \myo motors also play a crucial role for the mechanics and
adhesion of non-muscle tissue cells.  Cytoskeletal \myo assembles into
bipolar minifilaments consisting of $10$-$30$ motors
\cite{Verkhovsky1993}. They interact with an actin cytoskeleton which is
much less ordered than in muscle, mainly in the actin cortex as well as
in the contractile actin networks and bundles associated with cell
adhesion and migration \cite{Vicente-Manzanares2009}. Recently it has
been shown that the activity of \myo minifilaments contributes to the
sorting of actin filament orientation because of the asymmetric
elasticity of actin filaments \cite{Lenz2012, Murrell2012}. The \myo
based forces generated in the actin cytoskeleton are transmitted to the
extracellular environment via adhesion sites, which have been shown to
harbor different mechanosensitive processes \cite{Geiger2009,
uss:schw12b}. In particular, mature focal adhesions are often connected
to actin stress fibers consisting of parallel bundles of actin filaments
with alternating polarity enabling \myo minifilaments to contract the
bundles and thus mechanically load the adhesion sites. To apply these
forces effectively, the extracellular matrix underlying the adhesion sites
must not be too soft.
Therefore, cells are sensitive to the elasticity of the substrate and
adhere preferentially to stiffer substrates \cite{discher_tissue_2005}.
If the environment is much
stiffer than the cell, it essentially deforms itself and becomes
insensitive to the environmental stiffness \cite{uss:schw06a}. Therefore
cellular stiffness sets the scale for the sensitivity of
rigidity sensing \cite{solon_fibroblast_2007}. Due to
the complex interplay of many components in a cell, it is difficult to
identify the exact contribution of \myo to the rigidity response of
cells. One promising experimental route is the reconstruction of
\textit{in vitro} systems of motors and filaments \cite{Surrey2001,
Mizuno2007, Bendix2008, SoaresESilva2011, thoresen_reconstitution_2011,
thoresen_thick_2013,Murrell2012,shah_symmetry_2014}, which in the future might
allow us to probe these relations in more quantitative detail.

With the focus on the description of large assemblies of \myo motors in
the muscle sarcomere, theoretical progress has been made mainly through
mean-field models \cite{Huxley1957, Leibler1993, Vilfan1999} or computer
simulations \cite{Duke1999, Duke2000,gunther_spontaneous_2007}.
For ensembles consisting of a
large number of motors, details about internal motor states are less
important and experimentally accessible. Instead, collective quantities
such as velocity, walk length and number of bound motor are of large
interest. For example, generic two-state ratchet models have been used
to study the behavior of mechanically coupled motors
\cite{Juelicher1997, Placais2009, guerin_coordination_2010}. Here we aim
at understanding minifilaments with few \myo molecules for which
molecular details and stochastic effects are expected to be more
important. In this context, cross-bridge models are appropriate,
which have been studied before mainly with computer simulations
\cite{Duke1999, Duke2000, Walcott2012}. However, this approach is
numerically costly, in particular for extensions to systems with
multiple minifilaments. Recently the parallel cluster model (\pcm) based
on the cross-bridge cycle has been introduced as an efficient yet
detailed model for stochastic effects in small \myo ensembles
\cite{Erdmann2012, Erdmann2013}.

In this manuscript, we extend the \pcm to \myo minifilaments by modeling
them as two ensembles of \myo motors working against each other by
walking along two actin tracks with opposing polarity. This situation
can be considered as a tug-of-war of the two ensembles of non-processive
motors, in analogy to a tug-of-war of processive
motors \cite{Muller2008, grill_theory_2005}. In contrast to those
studies, however, we do not use a phenomenological force-velocity relation,
but rather a cross-bridge model to explicitly include the
molecular details of the motor cycle of \myo. In particular, we account
for the catch bond character of \myo unbinding (dissociation 
rate decreases under load, in contrast to the classical case of
a slip bond) and for the detailed kinetics of the power stroke.
From our model definition, it becomes clear that the mechanical
situation in bipolar \myo minifilaments is very complex, with an
effective spring constant that depends on internal mechanics,
external mechanics and the exact state of the motor ensembles.
Our main result is that \myo minifilaments show a kind of
mechanosensitivity that is reminiscent of the way cells respond to
environmental stiffness. We show that this effect not only results from
the molecular catch-bonding property, but also from the inability to
perform the power stroke in a stiff environment with sufficiently large force.
We also find that catch-bonding of \myo on stiff substrates leads to
frequent switches of direction of the ensemble movement and therefore to
an effective diffusion constant which decreases with increasing ensemble
size, in marked contrast to a tug-of-war of processive motors with slip bonds.

\section{Model}\label{scn:Model}

In the parallel cluster model (\pcm), individual \myo motors are described by a cross-bridge
model with three discrete states and stochastic transitions between them
\cite{Erdmann2012,Erdmann2013}. Here we generalize this model for \myo
minifilaments and discuss it with
parameter values originally introduced for modeling skeletal muscle
\cite{Duke2000,Vilfan2003}.  For cytoskeletal \myo, these values
depend on the exact isoform one is considering.
The parameter values used here result in a duty ratio of $0.33$, which 
lies in the range of duty ratios reported for cytoskeletal \myo B
\cite{kovacs_functional_2003,rosenfeld_myosin_2003}.
As shown schematically in
\fig{Setup}(a), a motor comprises three mechanical elements. The motor
head binds to the substrate and contains the \atp hydrolysis site, which
binds \atp or its hydrolysis products \adp and \pho. The rigid lever arm
is hinged to the motor head and alternates between stretched and primed
conformation, thus amplifying conformational changes in the motor head.
The linear elastic neck linker with spring constant $\kl=2.5 \pN\nm^{-1}$ anchors the
lever arm to the rigid motor filament. In the unbound (\ub) state, the
motor head is loaded with \adp and \pho and the lever arm is primed. The
motor reversibly transitions to the weakly-bound (\wb) state with
on-rate $\kne \simeq 40 \hz$ and off-rate $\ken \simeq 2\hz$. With
release of \pho, the lever arm swings to the stretched conformation and
the motor enters the post-power-stroke (\pps) state. The power stroke is
reversible with forward rate $\kez \simeq 10^3 \hz$ and reverse rate
$\kze \simeq \kez$ but is driven towards the \pps state by the free
energy bias $\Epp \simeq -60 \pN\nm$. This energy is stored in the
primed conformation of the lever arm as part of the energy released in
\atp hydrolysis.  Replacing \adp by \atp, unbinding and \atp hydrolysis
brings \myo from the \pps to the \ub state, thus completing the motor
cycle. These events are subsumed in a single reaction with unloaded rate $\kzn^0
\simeq 80\hz$ which is assumed to be irreversible due of the hydrolysis
of \atp. \Myo dynamics is characterized by the load dependence of power
stroke and unbinding from \pps state: the power stroke moves the lever
arm by the power-stroke length $d \simeq 8\nm$; unbinding from the \pps
state requires an additional movement of the lever arm by a distance
$\delta \simeq 0.3 \nm$. Thus, both reactions become slower under load.
The reduced rate of unbinding under load makes the \pps state of \myo a
catch bond rather than a slip bond in the range of forces considered here.

\FIG{Setup}{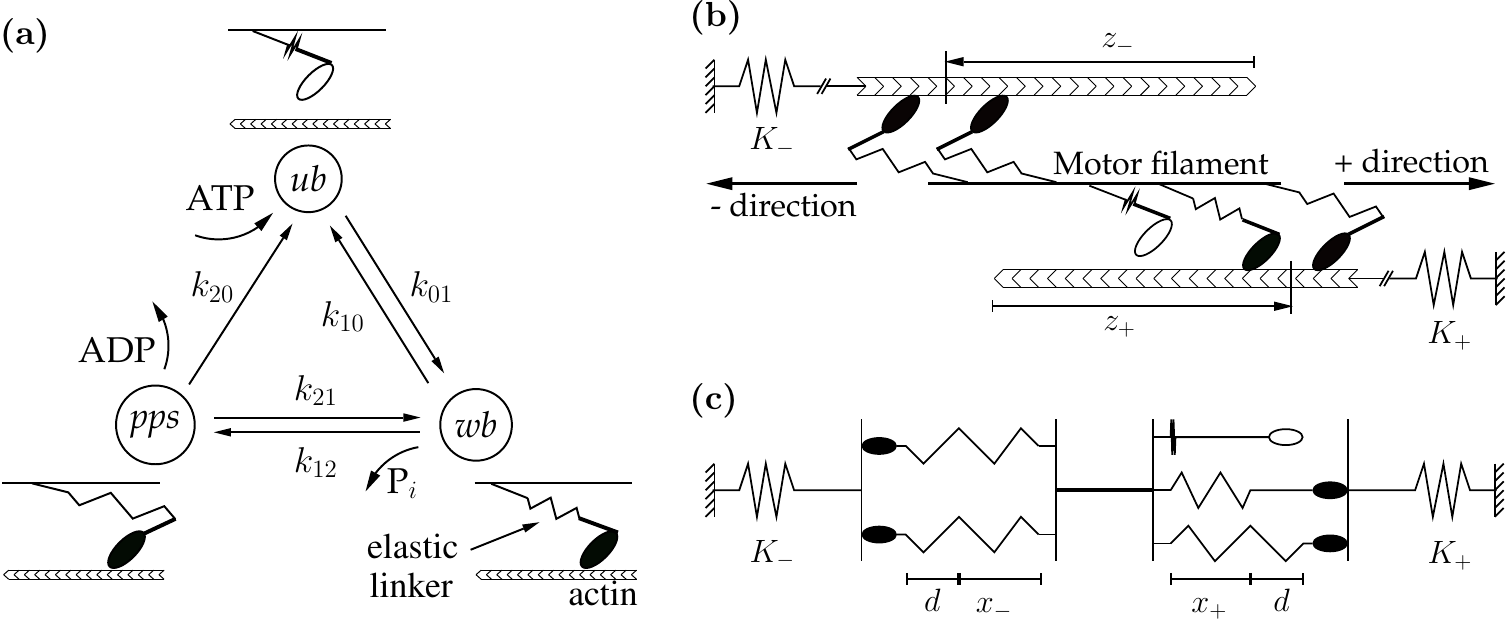}{(a) Mechanical setup
and hydrolysis cycle of \myo. In the unbound (\ub) state, the lever arm
is primed and the neck linker has vanishing strain. In the weakly-bound
(\wb) state the motor head is bound to actin. The lever arm is primed
but the neck linker generally has non-zero strain. In the
post-power-stroke (\pps) state the lever arm is stretched. Unbinding
from \pps state is the only irreversible transition because of the
hydrolysis of \atp. (b) Bipolar minifilament with $\Np = 3$ motors in
the \pe ensemble moving to the right and $\Nm = 2$ motors in the \me
ensemble moving to the left. The state of the \pme ensemble is described
by the number $\ipm$ of bound motors and the number $\jpm$ of motors in
the post-power-stroke state. The configuration in (b) corresponds to
$(\yp) = (2,1)$ in the \pe ensemble and $(\ym) = (2,2)$ in the \me
ensemble. The displacement of the ensembles on the actin filaments is
denoted by $\zp$ and $\zm$. Bound motors within each ensemble are
arranged in parallel; the two ensembles are arranged in series. (c) The
parallel cluster model (\pcm) applied to a bipolar minifilament treats
the motor ensembles as two adhesions clusters of parallel bonds coupled
in series with external springs. All bound motors in equivalent
mechano-chemical states have the same strain. The strain of weakly-bound
(\wb) motors is denoted by $\xp$ and $\xm$.  The power stroke shortens
the rest length of the neck linker by the power-stroke length $d$ so
that post-power-stroke (\pps) motors have the strain $\xpm + d$.}

The arrangement of \myo motors in a bipolar minifilament is depicted
schematically in \fig{Setup}(b). The minifilament consists of two
ensembles of motors working in opposite direction. The motors are
anchored to the rigid motor filament joining the two ensembles. Within
each ensemble, motors are arranged in parallel whereas the two ensembles
operate in series. The total number of motors in the ensemble working in
\pe direction (towards the right in \fig{Setup}(b)) is denoted by $\Np$;
the number of motors in the ensemble working in \me direction (towards
the left in \fig{Setup}(b)) is denoted by $\Nm$. The ensembles move on
actin filaments of opposite polarity which are attached to linear
elastic elements with spring constant $\Kp$ and $\Km$ to represent the
effective elasticity of the environment. The opposite polarity allows
the minifilament to slide the actin filaments relative to each other and
thereby to stretch the springs.  The first approximation of the \pcm is
to assume that motors in equivalent mechano-chemical states exert equal
forces (equal load sharing). This mean-field approximation is justified
by the small duty ratio of non-processive motors. The elongation
or strain of \wb motors in \pe and \me ensemble is denoted by $\xp$ and
$\xm$, respectively, so that \wb motors exert the force $\kl \xpm$. The
strain is positive, when the neck linker is stretched against the moving
direction of the ensemble (inwards in \fig{Setup}(b)). With the
assumption of equal load sharing, motor ensembles are mechanically
equivalent to adhesion clusters of parallel bonds as depicted in
\fig{Setup}(c) \cite{Erdmann2004}. In contrast to the adhesion cluster,
however, the rest length of the bond is not fixed, but is reduced by a
length $d$ due to the power stroke. Thus, motors in the \pps state have
the strain $\xpm + d$ and exert the force $\kl (\xpm + d)$. The strain
of \wb as well as \pps motors is determined by the offset $\xpm$ between
the bound motor head on the substrate and the anchor in the motor
filament. The state $(\yp;\ym)$ of a minifilament is described by the
numbers $\ip \leq \Np$ and $\im \leq \Nm$ of bound motors and $\jp \leq
\ip$ and $\jm \leq \im$ of \pps motors in both ensembles. The number of
\ub motors is $\Npm - \ipm$ and that of \wb motors $\ipm - \jpm$. With
$\jpm$ \pps motors and $\ipm - \jpm$ \wb motors, the \pme ensemble
exerts the force $\fpm = \kl \left[(\ipm - \jpm)\xpm + \jpm (\xpm +
d)\right]$. To determine the offset of \wb and \pps motors, we assume
that a minifilament is always in mechanical equilibrium. For the
arrangement of motor ensembles and external springs in series as in
\fig{Setup}(c) this requires that the ensemble forces $\fpm$ balance the
force $\fext$ exerted by the external springs. The latter is $\fext =
\Kext \left[\zp + \zm - (\xp + \xm)\right]$ where $\Kext = \Kp \Km /
(\Kp + \Km)$ is the effective external spring constant and $\zpm$ the
displacement of the \pme ensemble from its origin on the actin filament.
The extension of the external springs is $\zp + \zm - (\xp + \xm)$
assuming they are relaxed when $\zp + \zm = 0$ and all bound motors are
in the \wb state with $\xp = \xm = 0$. Solving the balance of forces,
$\fpm = \fext$, yields $\xpm$ as function of minifilament state
$(\yp;\ym)$ and contraction $\zsum := \zp + \zm$,
\begin{equation}\label{eq:XPM}
\xpm = \frac{\Kext (\zsum\imp + d\jmp) - (\Kext + \kl\imp) d\jpm}{\Kext (\ipm + \imp) + \kl\ipm\imp}\,.
\end{equation}
The force $\fext = \fpm$ as function of minifilament state and
contraction then reads
\begin{equation}\label{eq:Force}
\fext = \Ktot \left[\zsum + d\left(\frac{\jp}{\ip} + \frac{\jm}{\im}\right)\right]\,,
\end{equation}
where the total spring constant is defined as  
\begin{equation}\label{eq:Ktot}
\Ktot = \frac{\Kext\kl\ip\im}{\Kext(\ip+\im) + \kl\ip\im}
\end{equation}
and therefore varies dynamically, in contrast to $\Kext$. When all bound
motors in the \pe ensemble are in the \wb state ($\jp = 0$), the offset
is positive, $\xp > 0$, and the neck linkers are stretched against the
\pe direction. For growing contraction $\zsum \geq 0$, the offset $\xp$
increases because $\fext$ increases. A growing number $\jm$ of \pps
motors in the \me ensemble increases $\xp$ further, because transitions
to the \pps state shorten the minifilament and increase $\fext$. On the
other hand, $\xp$ decreases and can become negative for a growing number
$\jp$ of \pps motors in the \pe ensemble, although $\fext$ increases
further. For $\xp < 0$, \wb motors contribute to the external load which
is carried by the \pps motors whose strain is always positive, $\xp + d
> 0$.

The assumption of equal load sharing of equivalent motors defines a four
dimensional network of minifilament states $(\yp;\ym)$. The second
approximation of the \pcm reduces this network further by assuming that
bound states are in local thermal equilibrium (\lte). This is justified
by the strong separation of time scales between fast power-stroke and
slow binding kinetics \cite{Vilfan2003, Erdmann2012, Erdmann2013}. In
\lte, the conditional probability that $(\J)$ motors are in the \pps
state when $(\I)$ motors are bound is the Boltzmann distribution
\begin{equation}\label{eq:PSProb}
p(\I|\J) = \exp\left(-E/\kbT\right) / Z_{\I}\,,
\end{equation}
where $Z_{\I} = \sum_{\jp=0}^{\ip} \sum_{\jm=0}^{\im} \exp\left(-E /
\kbT \right)$ is the partition sum. The energy $E = \Eext + \Eelp +
\Eelm + (\jp + \jm) \Epp$ of a minifilament is the sum of the elastic
energy in the external springs, $\Eext = \Kext \left[\zsum - (\xp +
\xm)\right]^2 / 2$, the elastic energy $\Eelp + \Eelm$ in the neck
linkers, where $\Eelpm = \kl \left[(\ipm - \jpm) \xpm^2 + \jpm (\xpm +
d)^2\right] / 2$, and the free energy bias $\Epp < 0$ towards the \pps
state. Inserting \eq{XPM} for $\xpm$ yields $E$ as function of
minifilament state $(\yp;\ym)$ and contraction $\zsum$,
\begin{equation}\label{eq:FreeEnergy}
E = \frac{\Ktot}{2} \left[\zsum + d \left(\frac{\jp}{\ip} + \frac{\jm}{\im}\right) \right]^2 
  + \frac{\kl d^2}{2} \left[\frac{\jp\left(\ip-\jp\right)}{\ip} + \frac{\jm\left(\im-\jm\right)}{\im} \right]
  + \left(\jp + \jm\right) \Epp\,.
\end{equation}
The elastic energy is split into two contributions: the first is due to
overall stretching of external springs and neck linkers. For $\zsum \geq
0$, it increases with increasing contraction and number of bound motors
in \pps state. The second contribution is due to internal tension caused
by motors in different bound states. It vanishes when all bound motors
in an ensemble are either in \wb state or in \pps state, that is, for
$\jpm = 0$ or $\jpm = \ipm$, and is positive for intermediate states
with $0 < \jpm < \ipm$.

The \lte assumption for the bound states leaves the numbers $\ip$ and
$\im$ of bound motors as the only remaining variables. Binding and
unbinding changes the state $(\I)$ by $\ip \to \ip \pm 1$ and $\im \to
\im \pm 1$. Binding proceeds only to the \wb state and with constant
on-rate $\kne = \const$. Because $\Npm - \ipm$ motors can bind
independently the effective forward rate for the transition $\ipm \to
\ipm + 1$ in the \pme ensemble is
\begin{equation}\label{eq:ForwardRate}
\gpm(\ipm) = \left(\Npm - \ipm\right) \kne\,.
\end{equation}
The forward rate is independent of $\jpm$ and depends only on $\ipm$ in
the respective ensemble. Unbinding of motors proceeds either  with
constant off-rate $\ken = \const$ from the \wb state or with load
dependent off-rate $\kzn(\yp;\ym)$ from the \pps state. Thus, the
reverse rate for the transition $\ipm \to \ipm - 1$ in state
$(\yp;\ym)$ is
\begin{equation}\label{eq:Unbinding_Rate_jp_jm}
\rpm(\yp; \ym) = \left(\ipm-\jpm\right)\ken + \jpm\kzn(\yp;\ym)\,.
\end{equation}
The effective reverse rate for transitions in state $(\I)$ is obtained
by averaging over $\jpm$ with the \lte distribution from \eq{PSProb},
\begin{equation}\label{eq:ReverseRate}
\rpm(\I) = \sum\limits_{\jp=0}^{\ip} \sum\limits_{\jm=0}^{\im} \rpm(\yp;\ym) p(\J|\I)\,.
\end{equation}
We use a Kramers type load dependence for the off-rate from the \pps
state, $\kzn(\yp;\ym) = \kzn^0 \exp\left(-\kl (\xpm + d) / \Fdelta
\right)$. The off-rate decreases exponentially with increasing load $\kl
(\xpm + d)$ on a motor to describe the catch bond character of the \pps
state, where $\Fdelta = \kbT/\delta \simeq 12.6 \pN$ sets the unbinding
force scale. Inserting effective forward and reverse rate, a two
dimensional master equation for the probability $\pipim{\ip}{\im}(t)$
that $(\I)$ motors are bound can be formulated as
\begin{equation}\label{eq:MasterEq}
\begin{split}
\dd{\pipim{\ip}{\im}}{t} 
   &= \rplu(\ip+1,\im) \pipim{\ip+1}{\im} + \gplu(\ip-1) \pipim{\ip-1}{\im}\\
	&+ \rmin(\ip,\im+1) \pipim{\ip}{\im+1} + \gmin(\im-1) \pipim{\ip}{\im-1}\\
	&- \left[\rplu(\ip,\im) + \rmin(\ip,\im) + \gplu(\ip)  + \gmin(\im) \right] \pipim{\ip}{\im}\,.
\end{split}
\end{equation}
The probability for a specific state $(\yp;\ym)$ is the product of the
coarse-grained probability distribution $\pipim{\ip}{\im}(t)$ with the
conditional \lte probability distribution $p(\J|\I)$. The master
equation cannot be separated in two one-dimensional equations because
the reverse rates $\rplu(\I)$ and $\rmin(\I)$ depend on $\ip$ and $\im$
in both ensembles.

Because the effective reverse rates $\rpm(\I)$ depend on the contraction
$\zsum$, the master equation for binding dynamics has to be solved
together with rules for the displacement of the ensembles upon binding
and unbinding \cite{Erdmann2013}. We define the position $\zpm$ of an
ensemble as the average position of bound motor heads on the substrate.
The position of the motor filament is then given by $\zpm - \xpm$, where
$\xpm$ is the offset of the motors. New motors are assumed to bind with
vanishing offset, thus shifting the ensemble position by $\Delta \zpm =
-\xpm / (\ipm+1)$. Note that the offset $\xpm$ is negative for ensembles
subjected to small forces and  with the majority of motors in the \pps
state. To implement the rules for ensemble movement with \eq{MasterEq}
the offset needs to be averaged with the \lte distribution. This gives
$\xpmbar = \sum_{\jp=0}^{\ip} \sum_{\jm=0}^{\im} \xpm p(\J|\I)$ in state
$(\I)$ and the position change $\Delta \zpmbar = -\xpmbar / (\ipm + 1)$
upon binding of a motor. Unbinding of a motor does not change ensemble
position because all bound motors are assumed to be at the same position
$\zpm$. Complete detachment of one ensemble relaxes the external springs
and the still attached ensemble moves freely with offset $\xmp = -d$.
Reattachment therefore places the ensemble at $\zpm = \zmp + \xmp$.
Because $\zp$ and $\zm$ are defined in opposite directions, the sum
$\zsum = \zp + \zm$ describes the contraction of the actin substrates.
With these definitions, the \pcm for \myo minifilaments is completely
defined.

\section{Results}\label{scn:Results}

\subsection{Transitions in the power-stroke probability}

The \lte distribution $p(\J|\I)$ is shaped by the three parts of the
minifilament energy $E$ in \eq{FreeEnergy}. The second part---elastic
energy due to internal tension of opposing motors in both ensembles---is
symmetric against exchanging $\jpm$ and $\ipm - \jpm$. It vanishes when
all bound motors are either in \pps state ($\jpm = \ipm$) or in \wb
state ($\jpm = 0$). Between $\jpm = 0$ and $\jpm = 1$ (or $\jpm = \ipm$
to $\jpm = \ipm-1$) the energy increases by $\Delta E \geq \kl d^2/4
\simeq 40 \pNpernm \simeq 10 \kbT$ so that the relative occupancy is
$\ee^{-\Delta E/\kbT} \leq 10^{-4}$. This implies that intermediate
states with $0 < \jpm < \ipm$ are hardly occupied and $p(\J|\I)$ is
close to a binary distribution, in which either none or all of the bound
motors in an ensemble perform the power stroke. Only the four states
$(\J) = (0,0)$, $(0,\im)$, $(\ip,0)$ and $(\I)$, which are local minima
of $E$, can be appreciably occupied. The third part of $E$ in
\eq{FreeEnergy} decreases by the gain $\Epp < 0$ of conformational
energy for each of the $\jp + \jm$ \pps motors. This conformational
energy bias is opposed by the first contribution from the elastic energy
to $E$ in \eq{FreeEnergy} which increases with $\jpm$. This elastic
energy bias increases with contraction $\zsum$ (if $\jpm \geq 1$ and
$\zsum \geq -d$) and total spring constant $\Ktot$ and eventually
exceeds the conformational bias towards the \pps state.

\FIG{PSProb}{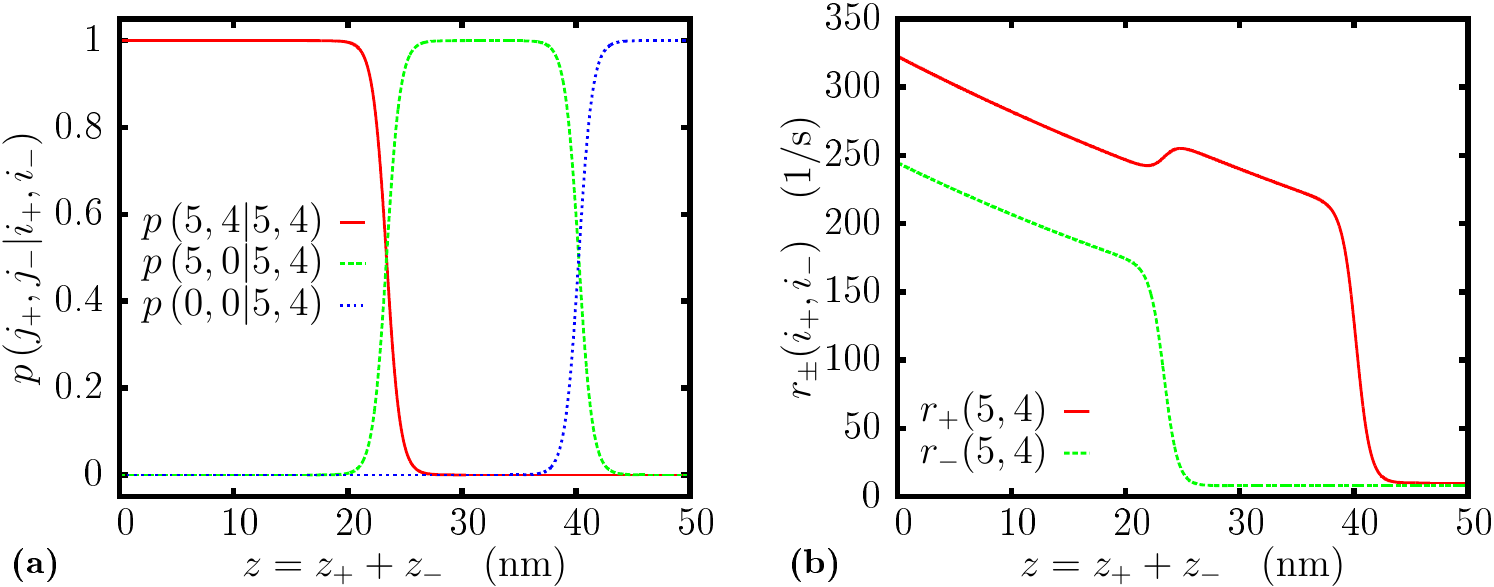}{(a) Power-stroke probability
$p(\J|\I)$ (see \eq{PSProb}) for a minifilament with $(\I) = (5,4)$
bound motors as function of contraction $\zsum = \zp + \zm$ for external
spring stiffness $\Kp = \Km = 2 \pNpernm$, that is, $\Kext = 1 \pNpernm$
and $\Ktot \simeq 0.85\pNpernm$. The power-stroke probability is shown
for $(\J) = (\I)$, $(\ip,0)$ and $(0,0)$; $p(0,\im|\I) = p(0,4|5,4)$ is
smaller than $10^{-7}$. (b) Effective reverse rates $\rplu(\I)$ and
$\rmin(\I)$ (see \eq{ReverseRate}) as function of $\zsum$ for the
minifilament from (a).}

\fig{PSProb}(a) shows the power-stroke probability $p(\J|\I)$ for a
minifilament with $(\I) = (5,4)$ bound motors as function of contraction
$\zsum$. For small $\zsum$, the gain of conformational energy in the
power stroke exceeds the increase of elastic energy and the minifilament
is in state $(\J) = (\I) = (5,4)$ with probability $p(\I|\J)  \simeq 1$.
At an intermediate value of $\zsum$, the minifilament switches to $(\J)
= (\ip,0) = (5,0)$ in a sharp transition. This means that above this
threshold, only the \pe ensemble with larger number of bound motors
($\ip > \im$) is able to perform the power stroke. Above a second
threshold, neither ensemble performs the power stroke and the
minifilament switches to $(\J) = (0,0)$. The plot confirms that the \lte
distribution can be almost neglected for intermediate states with $0 <
\jpm < \ipm$ and ensembles are said to be either in \pps or \wb state
according to the dominant state of the bound motors.

The thresholds for the transitions can be determined by comparing the
energy in the four states with $\jpm = 0$ or $\jpm = \ipm$. For $\ip >
\im$ the transition from $(\J) = (\I)$ to $(\ip,0)$ occurs at $\zsum
\geq -\im \Epp / (\Ktot d) - 3d/2$ and from $(\J) =(\ip,0)$ to $(0,0)$
at $\zsum \geq -\ip \Epp / (\Ktot d) - d/2$. In the symmetric case $\ip
= \im$, the states $(\J) = (\ip,0)$ and $(0,\im)$ are degenerate and the
minifilament occupies both with equal probability. It is important to
note that the thresholds contain an additional dependence on $(\I)$ via
$\Ktot$. For a given value of $\zsum$, fluctuations of $(\I)$ will
therefore induce transitions to the \wb state for small $\ipm$ and to
the \pps state for large $\ipm$.

\fig{PSProb}(b) plots the effective reverse rates $\rpm(\I)$ in \pe and
\me ensemble as function of $\zsum$ for the same minifilament setup as
in \fig{PSProb}(a). For small $\zsum$ the minifilament is in state $(\J)
= (\I)$ and both rates decrease exponentially with $\zsum$ due to catch
bonding of \pps motors. Note, however, that the relation between $\zsum$
and force depends on power-stroke probability: it is $\fext = \Ktot (z +
2d)$ for $(\J) = (\I)$, $\fext = \Ktot (z + d)$ for $(\J) = (\ip,0)$ and
$\fext = \Ktot z$ for $(\J) = (0,0)$.  At the transition to $(\J) =
(\ip,0)$, the bound motors in the \me ensemble can no longer perform the
power stroke and the effective reverse rate of the \me ensemble drops to
the value $\rmin(\ip,0) \simeq \im \ken$ determined by the small
off-rate $\ken \ll \kzn$ from of the \wb state. Because the
transition to the \wb state decreases the force on the $\ip = 5$ bound
motors in the \pe ensemble, the effective reverse rate $\rplu(\ip,0)$
increases during the transition. With the transition to $(\J) = (0,0)$
the bound motors in the \pe ensemble enter the \wb state and the
effective reverse rate drops to $\rplu(0,0) \simeq \ip \ken$.

\subsection{Stochastic trajectories}

The two dimensional master equation \eq{MasterEq} describes a
non-equilibrium process without detailed balance. Moreover, there is a
nonlinear feedback between minifilament displacement and binding
dynamics so that the master equation cannot be solved analytically.
Instead, we analyze minifilament dynamics numerically using the direct
method of the Gillespie algorithm. \fig{Trajectories} shows typical
stochastic trajectories of symmetric
minifilaments with varying size and external spring stiffness. The lower
panel of each plot shows the number of bound motors $(\I)$ in the two
ensembles. The upper panel plots the minifilament force $\fext = \fpm$
from \eq{Force} for a given $(\I)$ weighted with the appropriate
conditional probability over $(\J)$. Due to the binary nature of the
power-stroke probability, $(\J)$ is dominated by the states with $\jpm =
0$ or $\jpm = \ipm$. \fig{Trajectories}(a) shows a trajectory of a small
minifilament with $\Np = \Nm = 4$ motors for $\Kp = \Km = 0.2 \pNpernm$,
that is, $\Kext = 0.1 \pNpernm$. Due to the soft external springs, the
power stroke in both ensembles does not increase the load on the motors
appreciably so that the effective reverse rate remains close to its
large intrinsic value. Moreover, the threshold value of $\zsum$ for the
transition from \pps to \wb state is large and at least one of the
ensembles typically detaches before reaching the threshold. Therefore,
trajectories are characterized by frequent detachment of ensembles.

\SFIG{Trajectories}{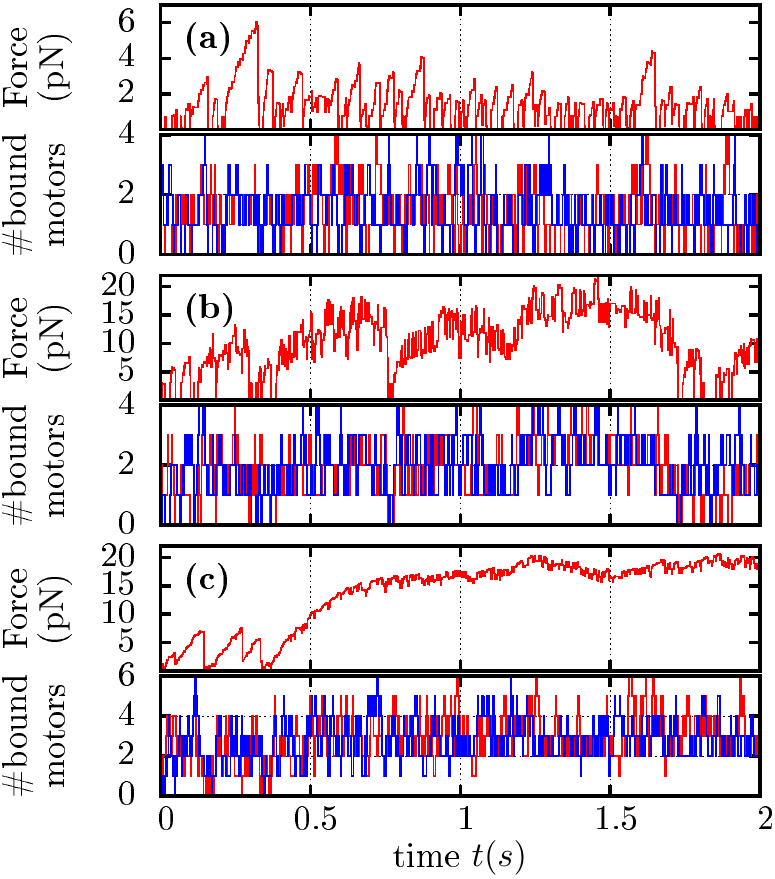}{Stochastic trajectories of
symmetric minifilaments ($\Np = \Nm$ and $\Kp = \Km$) with varying
ensemble size and varying external spring stiffnesses. In each plot,
the bottom panel shows the number of bound motors in \pe ensemble
($\ip$, red) and \me ensemble ($\im$, blue). The top panel shows the
force on the ensembles, that is, $\fext = \fpm$ (see \eq{Force})
weighted with $p(\J|\I)$. (a) $\Np = \Nm = 4$ and $\Kp = \Km = 0.2
\pNpernm$ ($\Kext = 0.1 \pNpernm$), (b) $\Np = \Nm = 4$ and $\Kp = \Km =
1.0 \pNpernm$ ($\Kext = 0.5 \pNpernm$) and (c) $\Np = \Nm = 6$ and $\Kp
= \Km = 0.2 \pNpernm$ ($\Kext = 0.1\pNpernm$).}

\fig{Trajectories}(b) shows a trajectory of a minifilament with the same
size as in \fig{Trajectories}(a) but for stiffer external springs with
$\Kp = \Km = 1.0 \pNpernm$ and $\Kext = 0.5 \pNpernm$. Detachment is
much less frequent than in \fig{Trajectories}(a) and the series of short
force peaks is no longer observed. Instead, initial attachment is
followed by a gradual increase of force towards a state with strongly
fluctuating but on average constant force. This is the result of two
effects. First, the force generated by the power stroke at $\zsum = 0$
decreases the off-rate of \pps motors appreciably so that the time to
detachment of the minifilament is increased. Second, the threshold for
the transition from \pps to \wb state is lowered and---in combination
with smaller off-rate---is more likely to be reached before detachment.
To stabilize attachment, it is sufficient that the power stroke cannot
be performed for $\ip = \im = 1$ so that the last motor in each ensemble
unbinds slowly from the \wb state. For $\ip = \im = 1$ the transition
from $(\J) = (\ip,0)$ to $(0,0)$ occurs at a force $\fext \simeq \Ktot
(z + d) \simeq 9 \pN$ below and $\fext \simeq \Ktot z \simeq 6 \pN$
above the transition. The fast fluctuations of force upon binding and
unbinding in \fig{Trajectories}(b) thus indicate transitions between
\pps and \wb state, while the slower variations are due to variations of
$\zsum$ following multiple binding events. Fluctuations to small values
of $\zsum$ allow both ensembles to perform the power stroke and the
increased reverse rate can lead to detachment of the minifilament.

\fig{Trajectories}(c) shows a trajectory of a larger minifilament with
$\Np = \Nm = 6$ motors for soft external springs $\Kext = 0.1\pNpernm$
as in \fig{Trajectories}(a). Initially, the miniflament detaches
repeatedly as observed for the smaller ensemble. Due to its larger size
and detachment time, the minifilament eventually reaches the threshold
$F \simeq 7.5 \pm 0.4 \pN$ above which the last motors in both ensembles
unbind from the \wb state. Although the minifilament is stalled for $\ip
= \im = 1$, it continues to builds up larger force because the typical
number of bound motors is larger. For forces above $F \simeq 16\pN$, the
trajectory of the number of bound motors shows that it becomes unlikely
to find a single bound motor. This force is large enough to keep the
last two bound motors in the \wb state with low unbinding rate and to
make detachment of any of the two ensembles unlikely.

The dynamics of force can be understood considering the sequence of \pps
to \wb transitions in \fig{PSProb}. For small $\zsum$, both ensembles
perform the power stroke and $\zsum$ increases quickly through the
activity of both ensembles. The increase of $\zsum$ is terminated by the
detachment of one of the ensembles. If the minifilament remains attached
sufficiently long, $\zsum$ reaches the threshold above which the
ensemble with smaller number of bound motors enters the \wb state. In
this case, the \pps ensemble moves forward upon binding of motors as
long as $\xpm < 0$. The \wb ensemble, on the other hand, will step
backward upon binding because $\xpm > 0$ without \pps motors. The
contraction $\zsum$ continues to increase as long as the forward
movement of the \pps ensemble is faster than the backward movement of
the \wb ensemble. As $\zsum$ reaches the threshold above which both
ensembles enter the \wb state, $\zsum$ can only decrease because $\xpm >
0$ for $\zsum > 0$ so that both ensembles step backwards upon binding of
motors. Since the threshold is reached first for small $\ipm$,
detachment of the minifilament becomes very unlikely. On the other hand,
for an increasing number of such states the minifilament enters an
isometric state in which contraction $\zsum$ and force $\fext$ fluctuate
with a constant average, because forward movement at large $\ipm$ is
balanced by backward stepping at small $\ipm$.

The trajectories in \fig{Trajectories} are for symmetric minifilaments
with the same number $\Np = \Nm$ of motors in \pe and \me ensemble and
equal external spring constants, $\Kp = \Km$. Differences of $\Kp \neq
\Km$ do not affect results, because $\Kpm$ enters the dynamic
description only via the effective external spring constant $\Kext$.
Differences of ensemble sizes, $\Np \neq \Nm$, do affect minifilament
dynamics but trajectories are qualitatively similar as long as the
difference is not too large. Most importantly, asymmetric minifilaments
display a net movement in the direction of the larger ensemble. When
both ensembles are attached and perform the power stroke, fewer motors
are bound on average in the smaller ensemble. These are subject to
larger force so that position steps are smaller or even negative; catch
bonding of \pps motors reduces this difference. Moreover, the smaller
ensemble detaches more frequently allowing the larger ensemble to move
freely. Finally, the smaller ensemble is more likely to transition from
\pps to \wb state and form a passive anchor for the larger ensemble.
The frequency of detachment of asymmetric minifilaments and the
probability to reach large forces, though, is determined by the smaller
ensemble.

\subsection{Mechanosensitivity}

The stochastic trajectories reveal a switch in minifilament dynamics
from transient attachment without sustained force (see
\fig{Trajectories}(a)) to near permanent attachment (see
\fig{Trajectories}(b-c)) in response to increasing external spring
stiffness. To elucidate this mechanosensitivity further,
\fig{MeanForce}(a) plots the mean force $\Fbar$ obtained by averaging in
the steady state generated by symmetric minifilaments with varying
ensemble size $\Np = \Nm$ as function of external spring stiffness $\Kp
= \Km = 2\Kext$. \fig{MeanForce}(a) reveals a strongly nonlinear
increase of $\Fbar$ with $\Kpm$, which is caused by catch bonding of
\pps motors in combination with \pps to \wb transitions in the
power-stroke probability.

\FIG{MeanForce}{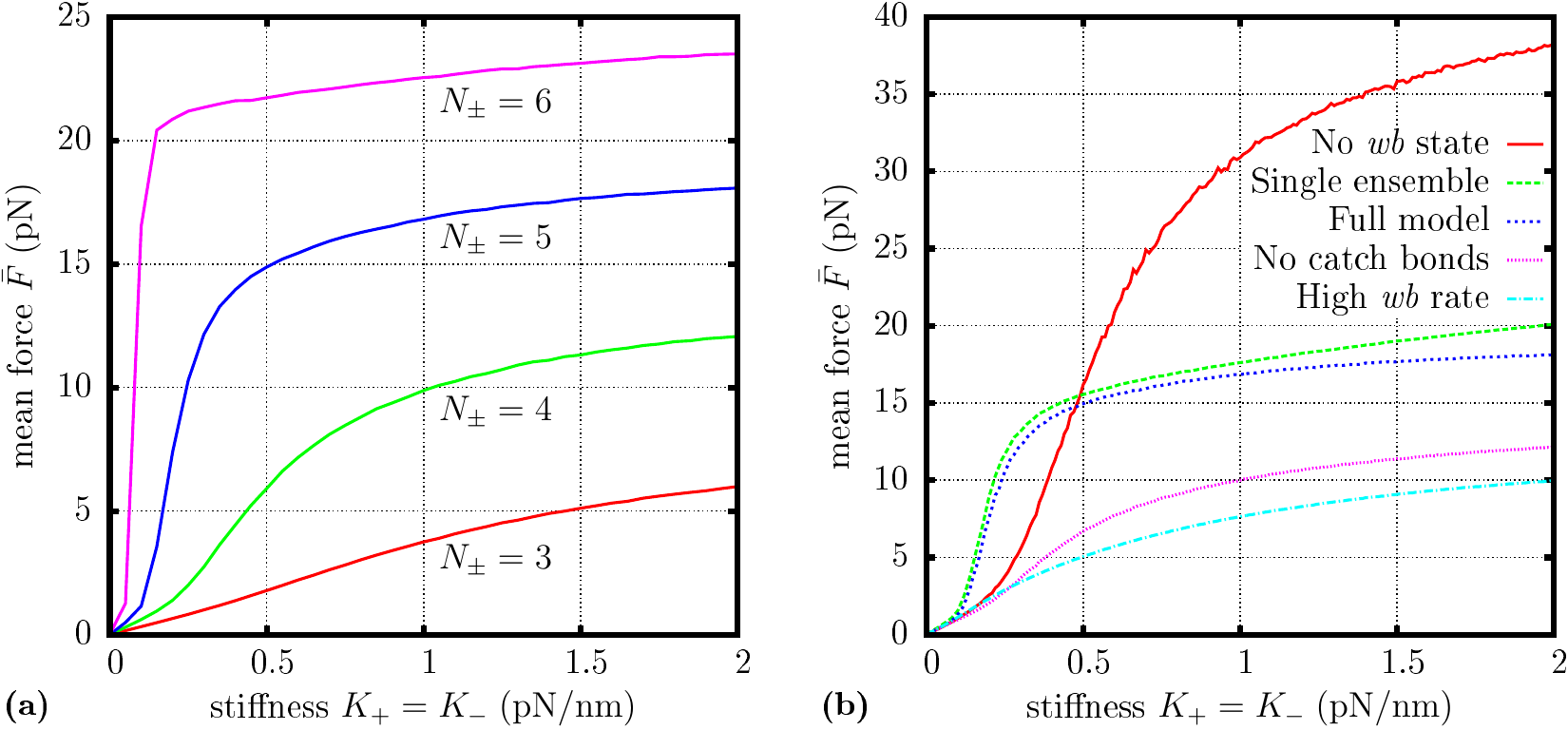}{Mean force $\Fbar$ generated by
symmetric minifilaments with varying number $\Np = \Nm$ of motors  as
function of external spring constant $\Kp = \Km = 2\Kext$. $\Fbar$ is
obtained by time averaging over stochastic trajectories. (a) Mean force $\Fbar$ for
$\Np = \Nm = 3$, $4$, $5$ and $6$. (b) Mean force $\Fbar$ for $\Np = \Nm
= 5$ for several model variants in comparison with the full model: $(i)$
no \wb state, that is, fixed power-stroke probability $p(\I|\I) = 1$,
$(ii)$ no catch bonding of \pps motors which unbind with constant
off-rate $\kzn \simeq 80\hz = \const$, $(iii)$ increased off-rate $\ken
=  80\hz = \const$ from the \wb state and $(iv)$ single \myo
ensemble working against an external spring with $\Np$ and $\Kp$ as for
the minifilament.}

For small stiffness, $\Fbar$ increases linearly with $\Kpm$. Here, the
contraction $\zsum$ reached before one of the ensembles detaches is on
the order of the power-stroke length $d$. For small values of $\Kext$,
the corresponding force is too small to increase the time to detachment
significantly, so that the typical contraction $\zsum$ hardly increases
with $\Kext$. In the limit of small $\Kext \ll \kl \ip \im / (\ip +
\im)$, it is $\Ktot \simeq \Kext$ so that the force $\fext = \Ktot (z +
2d) \sim \Kext$ is proportional to $\Kext$. The mean force along a
trajectory is proportional to the duty ratio of a minifilament, that is,
the fraction of time both ensembles are attached. The duty ratio is
hardly affected by small forces but increases with minifilament size so
that the slope in the linear regime increases with $\Npm$. Catch bonding
of \pps motors eventually increases detachment time and duty ratio so
that $\zsum$ increases beyond the force free case. The mutual positive
feedback between force and duty ratio leads to a rapid increase of
$\Fbar$ which is reinforced by transitions from \pps to \wb state. On
the other hand, these transitions limit the increase of force, because
\pps motors are required for forward movement of the ensembles and an
increase of $\zsum$. Once minifilaments are almost permanently attached,
$\Fbar$ increases slowly with further increasing $\Kpm$. This reflects
the increase of the average number of bound motors in the isometric
state, which is caused by increased force fluctuations for large $\Kext$
and the nonlinear dependence of the off-rate on force. The
increasing number of bound motors allows the minifilament to reach
larger values of $\zsum$ although the \pps to \wb transition occurs for
increasing values of $\ipm$. The linear regime is most prominent for
small $\Npm$ where the intermediate, super linear regime cannot be
discerned. With increasing $\Npm$, the linear regime shrinks and the
super linear growth in the intermediate regime approaches a step
increase of $\Fbar$, because detachment time depends sensitively on
changes of the off-rate for large ensembles \cite{Erdmann2013}.

\fig{MeanForce}(b) plots the mean force generated by minifilaments with
$\Np = \Nm = 5$ as function of $\Kpm$ for model variants in which
components of the \myo motor cycle are omitted in order to elucidate
their contribution to force generation. $(i)$ Without the transition
from \pps to \wb state (no \wb state), that is, for fixed power-stroke
probability $p(\I|\I) = 1$, the super linear increase of $\Fbar$ at
intermediate $\Kpm$ is comparable in steepness to the full model but is
shifted to larger values of $\Kpm$. Without stabilization through the
\pps to \wb transition, the increase of duty ratio and mean force is due
to catch bonding of \pps motors alone. On the other hand, $\Fbar$
reaches significantly larger values and continues to increase over the
whole range of $\Kpm$ because $\zsum$ is no longer limited by the
transition to the \wb state of both ensembles. Instead, $\Fbar$ is
limited by the stall force at which the offset $\xpm$ in both \pps
ensembles vanishes. Thus the transition of the power-stroke probability
is required to increase the sensitivity of the response and to generate
a switch-like behavior with a plateau at large forces. Without this
transition, the model would become unphysical because then the power
stroke would require more energy than provided by the \atp hydrolysis.
$(ii)$ Without catch bonding, that is, with constant off-rate $\kzn
\simeq 80 \hz = \const$ from the \pps state, the super linear increase
of $\Fbar$ at intermediate $\Kpm$ is present but occurs for larger
$\Kpm$ and is much weaker than in the full model. Also the overall level
of $\Fbar$ is strongly reduced. Trajectories show that minifilaments do
reach the transition from \pps to \wb state, but continue to detach
frequently. Thus, catch bonding of \pps motors does not only provide the
feedback between force and duty ratio needed for the steep increase of
$\Fbar$ at intermediate $\Kpm$, but also stabilizes the isometric state
by increasing the average number of bound motors. $(iii)$ With a large
off-rate $\ken \simeq 80 \hz = \const$ of \wb motors, transitions from
\pps to \wb state increase the effective reverse rates (see
\fig{PSProb}). Hence, \pps to \wb transitions induce detachment and
neutralize the effect of catch bonding. The level of $\Fbar$ is smaller
than in the case without catch bonding and a steep intermediate regime
is not observed. Thus a low off-rate from the \wb state is required to
conserve the large level of force in the isometric state and to generate
the observed switch-like response. $(iv)$ A single \myo ensemble working
against an external spring (representing one half of a minifilament),
generates slightly larger $\Fbar$ than minifilaments because the
frequency of detachment is reduced. Due to the larger total spring
constant, this difference increases as $\Kpm$ approaches $\kl$. For
very large $\Kpm$, on the other hand, the mean force generated by a
single ensemble collapses because the power stroke can no longer be
performed at $\zsum = 0$. This does not occur for minifilaments because
$\Ktot$ is limited by $\kl$. Thus the interplay between external and
internal mechanics is essential for the functioning of minifilaments.

\subsection{Ensemble movement on rigid substrates}

The contraction $\zsum$ is confined within a narrow range around a
stable fixed point for attached minifilaments. Fluctuations to large
$\zsum$ are limited by the transition of both ensembles to the \wb
state. Fluctuations to small $\zsum$ induce transitions to the \pps
state in both ensembles so that forward movement of both ensembles
increases $\zsum$ rapidly. The range of $\zsum$ narrows with increasing
$\Kext$. To facilitate analysis of this situation, we replace the
external springs by rigid anchorage.  The total spring constant reduces
to $\Ktot = \kl \ip \im / (\ip + \im)$ and the contraction $\zsum$ is
identical to the sum of the offset $\xpm$ in the two ensembles, $\zsum =
\zp + \zm = \xp + \xm$.  Due to the large stiffness of the neck linkers,
at most one ensembles can be in \pps state at $\zsum = 0$, while the
other is in \wb state.  The threshold for the transition to state $(\J)
= (0,0)$, at which the minifilament is stalled, is small and reached
within few binding steps.  As a consequence a stationary state is
established quickly and the time dependent solution
$\pipim{\ip}{\im}(t)$ of the master equation \eq{MasterEq} can be
replaced by the stationary limit $\pipimss{\ip}{\im} =
\pipim{\ip}{\im}(t \to \infty)$. Results of stochastic simulations for $\xpm$ reveal two
clearly separated peaks at a negative and a positive value of $\xp$
(data not shown). The peak at negative $\xp$ corresponds to the \pe
ensemble in \pps state with force $\kl(\xp+d)$ per motor and the peak at
positive $\xp$ to the \pe ensemble in \wb state with force $\kl \xp$ per
motor.  The separation of the peaks equals the power-stroke length, $d
\simeq 8 \nm$, which is the expected difference of $\xpm$ for $\ip =
\im$ (see \eq{XPM}). Finite width and asymmetry of the observed peaks
are due to fluctuation of $\ipm$ and $\zsum$ for the case $\ip \neq
\im$. The same results applies to the offset $\xm$ in the \me ensemble.

Stochastic simulations confirm that the
ensemble with larger number of bound motors usually performs the power
stroke while the other ensemble is in \wb state (see \fig{PSProb}). For
$\ip = \im$, the minifilament is in state $(\J) = (\ip,0)$ or $(0,\im)$
with equal probability. Therefore, the power-stroke probability from
\eq{PSProb} can be replaced by
\begin{equation}\label{eq:pwidjApprox}
\pwidjArgs = \Theta(\ip-\im) \Kronecker{\ip}{\jp} \Kronecker{0  }{\jm} 
           + \Theta(\im-\ip) \Kronecker{0  }{\jp} \Kronecker{\im}{\jm}\,.
\end{equation}
where $\Theta(x)$ is the Heaviside step function with $\Theta(0) = 1/2$
and $\Kronecker{i}{j}$ the Kronecker delta. As a consequence of the
mechanical coupling of ensembles and the mechanosensitivity of \myo, the
numbers $\ip$ and $\im$ of bound motors in \pe and \me ensemble are
synchronized. The correlation increases with $\Npm$ but is reduced for
minifilaments with soft external springs.  Synchronization is due to the
transition to the \wb state of the ensemble with fewer bound motors. The
small off-rate of \wb relative to \pps motors (see \fig{PSProb}(b))
tends to equalize the number of bound motors. Soft external springs
weaken the mechanical coupling of ensembles so that motors in one
ensemble are less sensitive to variations of the number of bound motors
in the other. Because minifilaments move in the direction of the \pps
ensemble, synchronization of the number of bound motors tends to reverse
the direction of motion of minifilaments and prevents long, persistent
runs. This is different from a tug-of-war of processive motors, which
are usually described as slip bonds which favor large differences of the
number of bound motors.

To derive an approximation for the stationary distribution
$\pipimss{\ip}{\im}$ we replace the continuous distribution of $\xpm$ by
a discrete one with two $\delta$-peaks. Assuming constant contraction
$\zsum$ during a typical binding and unbinding cycle through states
$(\I) = (i+1,i) \to (i+2,i) \to (i+2,i+1) \to (i+2,i) \to (i+1,i)$
allows to estimate the negative offset in the \pps ensemble as $\xneg =
-(d/2) (i + 1/2)(i+2)^2/(i+1)^3 < 0$. The offset in the \wb ensemble is
$\xneg + d > 0$. For $i = \Npm / 2$, excellent agreement of this
constant strain approximation with stochastic simulations is observed.
The load dependent off-rate of \pps motors becomes independent of
$(\I)$. The strain of \pps motors in the leading ensemble is $\xneg + d$
and their off-rate $\kzns = \kzn^0 \exp\left(-\kl (\xneg + d) /
\Fdelta\right)$. Together with the binary approximation of
\eq{pwidjApprox} for the power-stroke probability, the off-rate of
motors in the \pe ensemble (see \eq{ReverseRate}) reduces to
\begin{equation}
\rplu(\I) = \ip \left[ \Theta(\ip-\im) \kzns + \Theta(\im-\ip) \ken + (-\kzns +\kzn^0) \Kronecker{\im}{0} \right]\,.
\end{equation}
The first term describes unbinding from the \pps state, the second from
the \wb state and the third is for vanishing force in the case of a
minifilament with detached \me ensemble. An analogous expression holds
for $\rmin(\I)$.

\FIG{Probability}{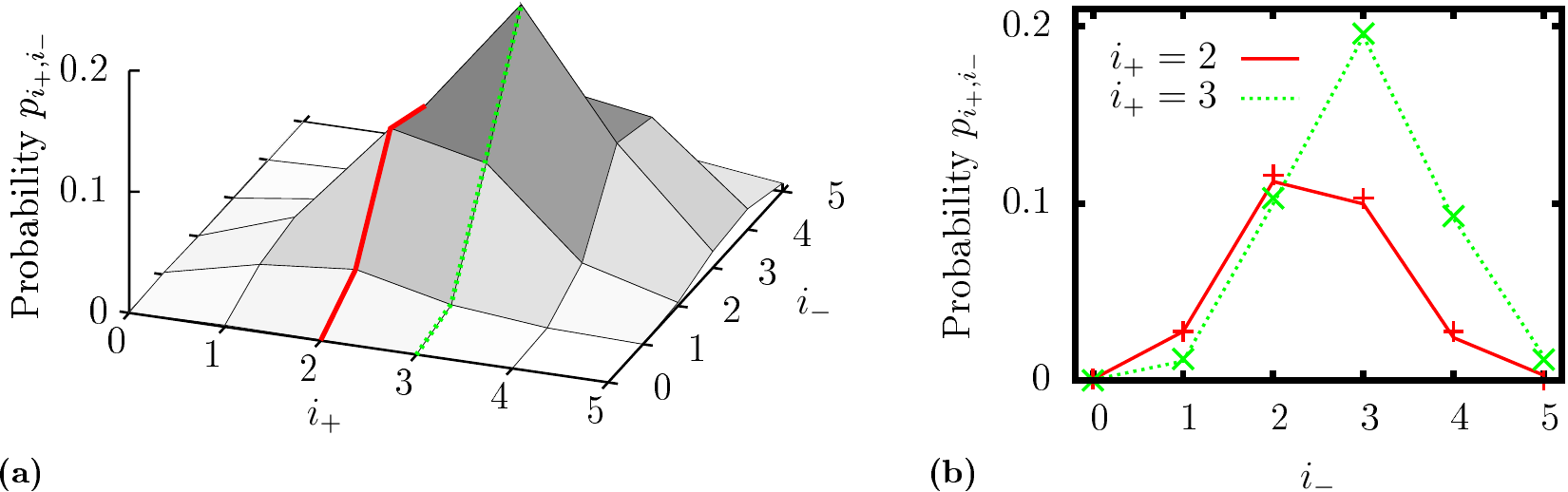}{Steady state distribution
$\pipimss{\ip}{\im}$ for symmetric minifilaments with $\Np = \Nm = 5$.
(a) Results of stochastic simulations for $\pipimss{\ip}{\im}$ as
function of $\ip$ and $\im$. Solid and dashed line indicate slices of
$\pipimss{\ip}{\im}$ for constant $\ip = 2$ and $\ip = 3$. (b)
Comparison of simulation results (lines) and analytical results
(symbols) from the constant offset approximation for
$\pipimss{\ip}{\im}$ as function of $\im$ for constant $\ip = 2$ and
$\ip = 3$.}

The constant offset approximation yields a two dimensional network of
states with constant transition rates per motor. Analytical solutions
for the stationary distribution are found by solving the corresponding
linear system of equations. \fig{Probability}(a) shows numerical results
for $\pipimss{\ip}{\im}$ from the exact model for a minifilament in a
tug-of-war with $\Np = \Nm = 5$. The distribution is symmetric with
respect to exchanging $\ip$ and $\im$ and strongly peaked at $\ip = \im
= 3$. It is centered along the diagonal with $\ip \simeq \im$ which
expresses the effect of synchronization of the ensembles.
\fig{Probability}(b) shows the stationary probability for fixed $\ip$ as
function of $\im$ and compares the numerical solution to the analytical
solution obtained via the constant offset approximation. Considering the
approximations entering the analytical solution the excellent agreement
is quite remarkable.

\SFIG{Diffusion}{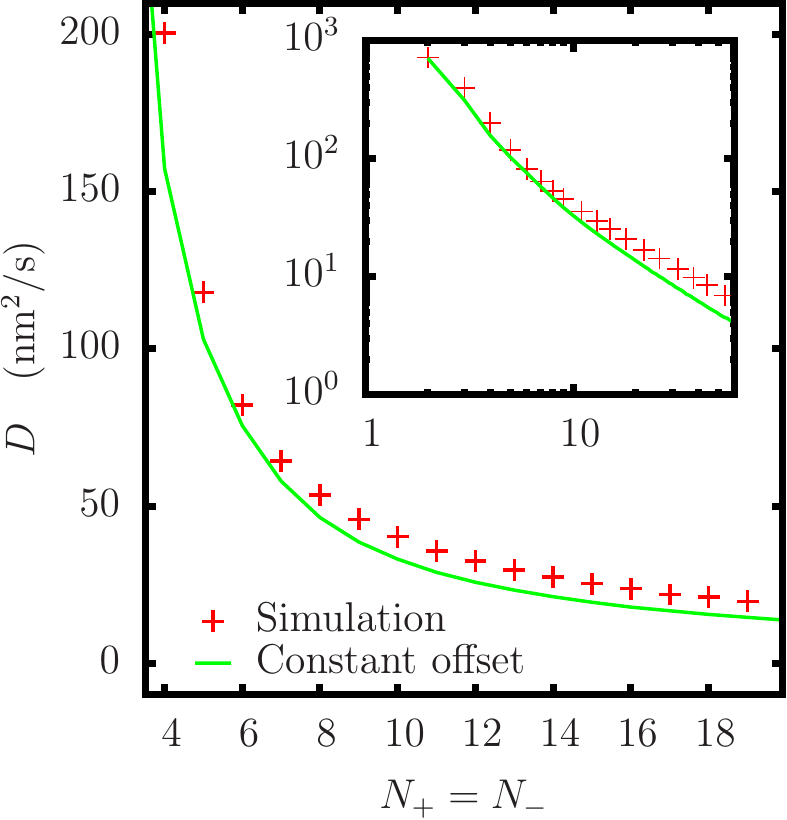}{Diffusion constant $D$ of a
minifilament as function of ensemble size $\Np = \Nm$ for symmetric
minifilaments. Numerical results (symbols) are compared to results from
the constant strain approximation (see \eq{DiffusionConstant}). 
The inset shows a logarithmic plot of the data for a wider range of $\Np$.}

The absolute position of the
minifilament can be defined as the mean position $\zabs = (\zp - \zm)/2$
of \pe and \me ensemble. Symmetric minifilaments perform an unbounded
random walk with vanishing mean which is characterized by the diffusion
coefficient $D$. Due to the limited range of the contraction $\zsum$,
the diffusion coefficient of $\zabs$ will be identical to that of $\zp$
and $\zm$ in the long time limit, $D \approx D_{+} = D_{-}$. We
calculate $D_{+}$ via the limit of the second jump moment per time in
the limit for a vanishing time step as $2 D_{+} = d_{+}$
\cite{Kampen2007}. For $\ip > 0$, binding in the \pe ensemble yields
$d_{+}^{\rm bound} = \gplu(\ip) \Delta \zp^2$ where $\Delta \zp =- \xp /
(\ip + 1)$. Detachment of the \pe ensemble releases the strain of both
ensembles and yields $d_{+}^{\rm detach} = \rplu(\I) (\xp+\xm)^2$ for
$\ip = 1$. Movement of the detached \pe ensemble through stepping of the
\me ensemble contributes by $d_{+}^{\rm drag} = \gmin(\im) \Delta \zm^2$
for $\ip = 0$ and $\im > 0$. The average diffusion constant of the
minifilament position $\zabs$ is
\begin{equation}\label{eq:DiffusionConstant}
D = D_{+} = \frac{1}{2} \sum\limits_{\ip=0}^{\Np} \sum\limits_{\im=0}^{\Nm} \left[d_{+}^{\rm bound} + d_{+}^{\rm detach} + d_{+}^{\rm drag}\right] \pipimss{\ip}{\im}\,.
\end{equation}
Within the constant offset approximation, the offset is $\xneg$ and
$\xpos = \im (\xneg + d)/\ip$ when both ensembles are attached and $-d$
if one of the ensembles is detached. Using the approximate expression
for $\pipimss{\ip}{\im}$ from \eq{pwidjApprox}, the contributions to the
diffusion constant reduce to
\begin{align}
 d_{+}^{\rm bound}  &= \frac{\gplu(\ip)}{(\ip+1)^2} \left[\Theta(\ip-\im) \xneg^2 + \Theta(\im-\ip) \xpos^2 \right] \quad\text{for}\quad \ip > 0\\
 d_{+}^{\rm detach} &= (\xneg+\xpos)^2 \left[\kzns \Theta(\ip-\im) + \ken \Theta(\im-\ip) \right]\quad\text{for}\quad \ip = 1\\
 d_{+}^{\rm drag}   &= \gmin(\im) \frac{d^2}{(\im+1)^2} \quad\text{for}\quad \ip = 0, \im > 0\,.
\end{align}
\fig{Diffusion} plots the diffusion constant $D$ of a minifilament as
function of ensemble size $\Np = \Nm$. The diffusion coefficient
decreases with increasing ensemble size. Simulation results show that
for large $\Npm > 8$, where detachment of ensembles is negligible, the
diffusion coefficient scales as the inverse of ensemble size, $D \sim
\Npm^{-1}$. For smaller $\Npm$, ensemble detachment leads to a stronger
dependence on $\Npm$. In the approximation of \eq{DiffusionConstant},
only $d_{+}^{\rm bound}$ contributes to $D$ for $\Npm > 8$. For a given
$i$, the squared step $\Delta \zpm^2$ scales as $1/(i+1)^2$ and the
forward rate as $\Npm - i$. Using $i \simeq \Npm/2$, this yields the
scaling expression $D \sim \left(\Npm (1 - 2 / \Npm^2) \right)^{-1}$
which fits well to the observed dependence $D \sim 1/\Npm^{1.12}$. The
deviations between simulation and analytical results are probably due to
correlations in the fluctuations of the offset $x$. The decrease of $D$
is in contrast to the dependence of $D$ on ensemble size in a tug-of-war
of processive motors. Processive motors are usually described as slip
bonds so that the time to switch between directions increases
exponentially with ensembles size, which induces an exponential increase
of the diffusion coefficient \cite{Muller2010}. The synchronization of
the number of bound motors as a consequence of the catch bond character
of \myo leads to a weak dependence of the switching time on ensemble
size and prevents the increase of the diffusion coefficient.

\section{Discussion and Conclusion}\label{scn:Discussion}

Here we have introduced a new model for the stochastic dynamics of \myo
minifilaments. Analyzing the stochastic trajectories and the mean force
generation predicted by our model revealed a switch of the dynamic
behavior of the minifilaments in response to changes of the
environmental elasticity as a consequence of two independent
mechano-sensitive processes. First catch bonding of \pps motors depends directly on the load
on the motors. A gradual increase of load decreases the reverse rate and
increases the number of bound motors to stabilize ensemble attachment.
Second the transition from \pps to \wb state provides another type of catch
bonding because unbinding from \wb state is slower than from \pps state,
$\ken \ll \kzn$. There are two important differences: $(i)$ \pps to \wb
transitions reduce the reverse rates $\rpm(\I)$ abruptly and $(ii)$ the
transition is sensitive not only to minifilament load but also to the
elasticity of the environment. For single \myo ensembles, \pps to \wb
transitions only occur in elastic environments, but not for ensembles
working against constant external load \cite{Erdmann2012, Erdmann2013}.
In a miniflament, the elastic neck linkers provide an elastic
environment for the ensembles even with rigid anchorage. Therefore
our model demonstrates the importance of considering both the internal and
external mechanics when investigating the function of a \myo minifilament.

Both mechano-sensitive processes cooperate in stabilizing minifilament
attachment. While permanent attachment can be reached with catch bonding
of \pps motors alone, the transition from \pps to \wb state shifts the
switch to smaller external stiffness and steepens the transition. On the
other hand, the transition to the \wb state alone is not sufficient for
stable attachment. The two mechano-sensitive processes have contrasting
effects on minifilament dynamics. Catch bonding of \pps motors allows to
distribute load on a growing number of bound motors.  This increases the
velocity of an ensemble at a given load and increases the stall force.
In the context of sarcomeric and cytoskeletal \myo ensembles working
against a constant external load, it was shown that catch bonding of
\pps motors leads to the characteristic upward convex force-velocity
relation \cite{Duke1999, Duke2000, Erdmann2012, Erdmann2013}. For
minifilaments, catch bonding of \pps motors increases the mean force
exerted in an elastic environment. Transitions from \pps to \wb state,
on the other hand, limit the increase of force because \pps motors are
required for an increase of contraction, so that minifilaments adjust
themselves to an isometric state. In single \myo ensembles working
against a linear external load, the force generated by the ensembles
breaks down for very stiff external load because the power stroke can no
longer be performed at $\zsum = 0$. In minifilaments, the effective
stiffness is limited by the elasticity of the neck linkers and this
breakdown is prevented. Therefore, minifilaments in the isometric state
generate a mean level of force, which depends mainly on the number of
motors available for binding, but is robust to variations of elastic
properties of the environment. Transitions to the \wb state limit the
mean force as long as they occur before the stall force of an ensemble in
\pps state is reached.  This is the case for the parameters in our
model; for significantly smaller neck linker stiffness as used in Ref.\
\cite{Walcott2012} with comparable power-stroke length, the stall force
is reduced strongly and could be reached before the transition to the \wb
state. For the smaller stall force, however, catch bonding of \pps
motors would not be sufficient to decrease the reverse rate sufficiently
to achieve stable attachment.

The mechanosensitivity of \myo minifilaments has important implications
for the structure and function of acto-myosin networks. In the actin
cortex or reconstituted mixtures of actin with \myo minifilaments, the
apparent stiffness of an actin filament depends largely on the level of
crosslinking. Due to their mechanosensitivity, \myo minifilaments will
most efficiently form crosslinks with and exert force to filaments which
are already firmly linked to the network.  Loose filaments, on the other
hand, which do not contribute to overall network tension, could not be
integrated by \myo minifilaments. This could help to create
densely linked networks in which minifilaments provide active crosslinks
maintaining a constant level of tension, which is restored after
deformations. Stretching a network would initially reinforce attachment
and prevent rupture of the network. On longer time scales, the increased
tension would be released by the flow of the network as a consequence of
backward stepping of minifilaments. Compression, on the other hand,
would reduce the load on the minifilament crosslinks. Subsequent forward
movement of \myo ensembles then helps to restore the initial tension in
the network. Persistent compression, however, could lead to \myo
detachment and network disintegration. In an actin network of random
polarity, mechanosensitivity of \myo minifilaments in combination with
the asymmetric elasticity of actin filaments could lead to a selection
of filaments according to their orientation, because large forces can be
exerted when filaments are stretched, but small forces on filaments
buckling under compression \cite{Lenz2012, Murrell2012}. This could also
contribute to the observed contraction rather than expansion of actin
networks of random orientation. Unbalanced preferential attachment to
strongly linked filaments could cause aggregation of \myo minifilaments
and actin, as it is observed in reconstituted assays
\cite{SoaresESilva2011}. In adherent cells elasticity of the
extracellular matrix contributes to the effective stiffness experienced
by \myo minifilaments. Actin stress fibers tend to form and reinforce at
focal contacts on sufficiently stiff substrates \cite{Lo2000,
Geiger2001, Geiger2009}. Mechanosensitivity of \myo minifilaments
could contribute to discrimination of substrate stiffness because they are not able to
crosslink and to build up forces in soft environments. Although this
possible relation is rather speculative at this stage and the exact
role of myosin minifilaments in the rigidity response of cells is hard
to pin down, it is worth mentioning that \myo activity is an integral
part of mechanosensing, which is completely disrupted by inhibiting \myo
activity, e.g.\ with the pharmacological inhibitor blebbistatin \cite{discher_tissue_2005}.

According to our model, the dynamics of \myo minifilaments is strongly
determined by the existence of at least two load bearing bound states of
\myo motors. For small load, the free energy bias allows the motors to
perform the power stroke and complete the motor cycle; for large load,
however, the increase of the elastic energy exceeds the free energy bias
and the power stroke can no longer be performed. Because the free energy
bias is limited by the energy gained in \atp hydrolysis, this transition
is inevitable, independent of the parameters. The exact value of the
transition thresholds, however, depends on the stiffness $\kl$ of the
neck linkers. This also means that the value of $\kl$ determines the
range of external spring stiffness $\Kpm$ to which minifilaments are
sensitive. The values for $\kl$ used in the literature range over an
order of magnitude, $\kl = 0.3 \pNpernm \dots 3 \pNpernm$
\cite{Walcott2012, Duke1999, Duke2000, Chen2011}. Although the effect of
varying $\kl$ is reduced by combining small $\kl$ with large values of
the power-stroke length $d$, small values as in Ref.\ \cite{Walcott2012}
could lead to stalling of minifilaments before the transition of the
power-stroke probability is reached. Unless, however, the unbinding
force scale $\Fdelta$ is decreased as well, this stall force would not
be enough to achieve stable attachment. Another set of parameters
determining the dynamics of \myo minifilaments in our model are the on-
and off-rates of motors, which determine the duty ratio. The load-free
duty ratio ($\rho \simeq 0.33$) of \myo with the transition rates used
here is comparable to values reported for non-muscle \myo isoform B but
significantly larger than for non-muscle \myo isoform A or smooth muscle
\myo. Changes of the transition rates will not change the outcome of the
model qualitatively as long as the off-rate from the \wb state is small
compared to the off-rate from the \pps state under load. Smaller values
of the duty ratio could be compensated by larger ensemble sizes in order
to achieve the stabilization of minifilament attachment observed here.

We have further shown with an analytical treatment for the case of 
minifilaments on rigid substrates that the effective
diffusion constant decreases with ensemble size, in marked
contrast to the case of the tug-of-war of processive motors, where the
diffusion coefficient increases exponentially with the number of motors
\cite{Muller2010}. Processive motors are usually described as slip bonds
favoring strongly asymmetric states, in which one of the ensembles is
completely detached. With increasing number of motors, a tug-of-war of
processive motors generates persistent movement in either direction, with
exceedingly rare reversal of the direction of motion. This dependence of
the diffusion coefficient on ensemble size is in line with the function
of processive motors for transport, which becomes more efficient with
increasing number of motors. Non-processive \myo motors, on the other
hand, serve as active, force generating crosslinks which are not
required to move through the cytoskeleton. An experimental realization
of a tug-of-war of two ensembles of non-processive motors, e.g.\ using a
solid substrate covered by actin filaments with nematic order but random
polarity or bipolar actin constructs on a surface covered with \myo
motors, could be used to investigate the characteristic difference
between processive, slip-bond motors and non-processive, catch-bond
motors. In general, we envision that biomimetic assays provide rewarding
avenues to experimentally test our theoretical predictions.

\begin{acknowledgments}
The authors acknowledge support by the EU-program MEHTRICS.
USS is a member of the Heidelberg cluster of excellence CellNetworks.
\end{acknowledgments}

\bibliographystyle{apsrev}

\end{document}